\documentstyle[12pt]{article}
\begin{document}

\begin{titlepage}

\baselineskip=18 pt

\begin{flushright}
   SU-HEP-4241-531\\
   February 1993\\
\end{flushright}

                  \vspace{2 cm}

\begin{center}
  \Large\bf Comments on the Gribov Ambiguity\\

                  \vspace{2.5 cm}

\baselineskip=18pt
\large\sc Nikolaos Kalogeropoulos $^{\dagger \ \star}$\\
                   \medskip
\normalsize\sl Department of Physics\\ Syracuse University\\
               Syracuse, NY 13244-1130\\
\end{center}

                   \vspace{2 cm}

\begin{center}
\large\sc ABSTRACT\\
\end{center}

\rm We discuss the existence of Gribov ambiguities in $SU(m)\times U(1)$
    gauge theories over the $n-$spheres. We achieve our
    goal by showing that there is exactly one conjugacy class of groups of
    gauge transformations for the theories given above. This implies that
    these transformation groups are conjugate to the ones of the trivial
    $SU(m)\times U(1)$ fiber bundles over the
    $n-$spheres. By using properties of the space of maps $Map_{\ast}(S^n,G)$
    where $G$ is one of $U(1)$, $SU(m)$ we are able to determine the
    homotopy type of the groups of gauge transformations in terms of the
    homotopy groups of $G$. The non-triviality of these homotopy groups gives
    the desired result.\\

                   \vfill

\footnoterule
\noindent\footnotesize $^{\dagger}$ Research supported in part by the DOE grant
              DE-FG02-85ER40231\\
              $^{\star}$ e-mail \ \ nikolaos@suhep.phy.syr.edu
\end{titlepage}

\baselineskip=18pt  %%%%%%% Beginning of the main body of the paper %%%%%%%

                    \newpage

      \begin{center}
         \large\sc 1.\ \ Introduction\\
      \end{center}

\rm Despite their striking successes, non-Abelian gauge theories are far from
being well-understood. Most of the progress that has been done is within the
region of validity of perturbation theory. Very few things are known about
non-perturbative effects, by comparison. Although this lack of understanding is
also a problem for QED, it becomes a much greater subject of concern for
non-Abelian gauge theories, in particular for the ones that describe such
striking physical effects as confinement. The confinement problem is one of
enormous importance in QCD. It has become obvious, through extensive studies
over a number of years, that this is a genuinely non-perturbative effect.
Therefore, its solution requires us to go beyond perturbation theory, in a
regime which is virtually unexplored both physically and mathematically.\\

One of the many characteristics of the non-Abelian gauge theories is the
existence of Gribov ambiguities \cite{Gribov}. This is the fact that  the
Coulomb or Lorentz gauges (or more generally any covariant gauge) fails to
globally eliminate the spurious degrees of freedom of a theory defined over a
$3-$ or $4-$ sphere \cite{Singer}.
This effect does not appear when we use the axial gauge
(or more generally, algebraic gauges) but in that case we cannot compactify
our space or spacetime to a sphere.
 The physical implications of the
existence of the Gribov ambiguities are not known. Despite the early claims,
that this phenomenon could provide a solution to the confinement problem
\cite{Gribov}, \cite{Zwaz}, no
definitive proof has ever been presented, and it is not unreasonable to say
that the problem remains still open. It is worthwhile, therefore, to further
explore the meaning of the Gribov ambiguity both physically and mathematically
since it is one of the very few known non-perturbative effects of the
non-Abelian gauge theories.\\

In this paper we present some mathematical arguments for the existence of
Gribov copies of an $SU(m)\times U(1)$  gauge theories
over $n-$spheres. In section 2 we present the statement of the Gribov problem.
In section 3 we discuss the group of automorphisms (group of gauge
transformations)
of some principal fiber bundles and the associated conjugacy classes. In
section 4
we apply these methods to explicitly determine the homotopy type of the
automorphism groups of $SU(m)\times U(1)$ principal
fiber bundles over $n-$spheres in terms of the homotopy groups of $SU(m)$ and
$U(1)$ respectively. In section 5 we present our conclusions.\\

                            \vfill

                            \newpage

      \begin{center}
         \large\sc 2.\ \ The Gribov problem\\ %%%%%%%%%%%%%%%%%%%%%%%%%%%%%%
      \end{center}

\rm In classical electrodynamics on $R^4$, the condition $\partial
^{\mu}A_{\mu} = 0$ fixes the gauge completely. The Lorenz gauge completely
eliminates all the redundant degrees of freedom. Under a gauge transformation
 $$A_{\mu} \rightarrow A_{\mu}^\prime = A_{\mu} + \partial_{\mu}\Lambda$$
$A_{\mu}^\prime$ satisfies the Lorenz condition provided that
$\partial^{\mu}\partial_{\mu}\Lambda = 0$. In order to have a finite energy
configuration we impose the following boundary condition at infinity
$A_{\mu}\stackrel{r\rightarrow\infty}{\rightarrow}\frac{1}{r}$ (we also assume
that
$A_{\mu}$ is regular everywhere). This allows us to treat infinity as a point.
So, from the viewpoint of dynamics the space on which our theory is defined is
$S^4$ (one-point compactification of $R^4$). More generally, to assure
finiteness of the energy of the system we confine ourselves to a compact
submanifold of $R^4$ with boundary and we are imposing vanishing boundary
conditions for the fields. In that case due to the ellipticity of
$\partial^{\mu}\partial_{\mu}\Lambda = 0$ the only solution is $\Lambda=0$. We
see that by imposing the Lorenz gauge on a compact space with boundary we have
completely eliminated the spurious degrees of freedom. \\

The situation is entirely different in non-Abelian gauge theories. In that case
a gauge transformation is
  $$A_{\mu}^{a}\rightarrow A_{\mu}^{a}+D_{\mu}^{ac}\Lambda^{c}, \ \ \ \
            D_{\mu}^{ac} = \partial_{\mu}\delta^{ac} + gf^{abc}A_{\mu}^{c}$$
When we impose the Lorenz gauge condition on the new and the transformed
gauge potential we get
 $$\partial^{\mu}\partial_{\mu}\Lambda^{a} + gf^{abc}A_{\mu}^{b}\partial^{\mu}
                               \Lambda^{c} = 0$$
Gribov \cite{Gribov} proved that the above equation admits non-trivial regular
solutions $\Lambda^{c}$ for large values of $A_{\mu}^{a}$. This means that the
Lorenz condition does not eliminate all the spurious degrees of freedom in a
non-Abelian gauge theory defined over a compact space. There is no trouble
with this effect in case we are interested solely in perturbative effects. In
a non-perturbative treatment however this phenomenon becomes important.\\

To describe precisely, in a geometric way, the Gribov ambiguity let's consider
a
Riemannian manifold $M$ with Euclidean signature metric $g_{ab}$ and let $G$ be
a gauge group. Let $P$ be a principal fiber bundle with typical fiber $G$ over
$M$ and let $\pi : P\rightarrow M$ be the canonical projection of $P$ onto $M$.
Following \cite{AHS},\cite{MV} we introduce the associated to $P$ bundle of
groups
$AdP=P\times_{G}G$ where $G$ acts on $G$ through the adjoint action and the
associated bundle of Lie algebras $adP=P\times_{G}LieG$ where $LieG$ is the Lie
algebra of $G$ and the action of $G$ on $LieG$ is still the adjoint action. By
a connection (gauge potential) we mean a $LieG$ algebra valued
one-form on $P$. (Since we are not primarily interested in the functional
analytic aspects of the problem we assume from now on that all fields and maps
are $C^{\infty}$. A careful treatment when this is not the case is presented
in\cite{MV},\cite{NR}).
For any two such gauge potentials $A_{1}$ and $A_{2}$ their difference
$A_{1} - A_{2}$ is pulled back on $M$ using $\pi$. This means that
  $$A_{1} - A_{2} \in \Omega^{1}(M;adP) = \Gamma(T^{\ast}M\otimes adP)$$
Let us denote by $\cal A$ the space of gauge potentials. We see that it is an
affine space modelled over the vector space $\Omega^{1}(M;adP)$. By an
automorphism of $P$ we mean a diffeomorphism $f:P\rightarrow P$ which preserves
the right action of $G$ on $P$ i.e. $f(pg)\rightarrow f(p)g$, $p\in P$, $g\in
G$. The space of gauge transformations is the space of sections (with pointwise
multiplication) of $AdP$ and it has been proved to form a Lie group i.e.
                       $${\cal G} = \Gamma (AdP)$$
Let $f\in\cal G$ and $A\in\cal A$. Then we have the following action
                   $${\cal A}\ni A\mapsto f^{\ast}A\in\cal A$$
The problem is that this action is generally not free. This means that there
are connections $A$ such that $fA=A$ with $f\neq id$. Non-free actions are
the origin of orbifold-type singularities which are quite complicated to
handle. In order to simplify technically our treatment we do not consider the
space of connections but instead the space of framed connections. A framed
connection \cite{DK} in $P$ is a pair $(A, \phi)$ where $A$ is a connection on
$P$ and
$\phi$ is an isomorphism of $G-$spaces $\phi : G\rightarrow P_{x_0}$. The group
of gauge transformations acts naturally on the framed connections giving the
moduli space
\begin{equation}
\tilde{\cal A} =  ({\cal A} \times Hom(G, P_{x_0})) /\cal G
\end{equation}
An alternative way of thinking about $\tilde{\cal A}$ is to fix the framing
$\phi$ and define $\cal G_{\ast} \subset \cal G$ to be the isotropy group at
point $p_0$ i.e.
\begin{equation}
{\cal G}_{\ast} = \{ f\in{\cal G} : f(p_{0}) = 1,\ \ \ \ p_{0}\in P\}
\end{equation}
Then $\tilde{\cal A}=\cal A/\cal G_{\ast}$. The isotropy group at $A\in\cal A$
is $\Gamma_{A} = \{ f\in{\cal G} : fA=A\}$. Since $fA = A - (D_{A}f)f^{-1}$ we
see
that $\Gamma_{A} = \{ f\in{\cal G} : D_{A}f=0\}$.
Because the isotropy group at $A$ consists of covariantly constant gauge
transformations the subgroup $\cal G_{\ast}$ acts freely on $\cal A$. Actually
it has been proved that $\tilde{\cal A}$ is a Hilbert manifold
\cite{MV},\cite{NR}.
Physically the
point $x_{0}$ can be chosen to be infinity. Then $\cal G_{\ast}$ is the group
of gauge transformations that are identity at infinity. Finally, we note that
$\tilde{\cal A}$ is a principal $\cal G_{\ast}$ bundle over $\tilde{\cal
A}/\cal G_{\ast}$.\\

Gauge fixing means picking one connection from each orbit of $\tilde{\cal
A}/\cal G_{\ast}$. This corresponds to a section $s : \tilde{\cal A}/{\cal
G}_{\ast}\rightarrow \tilde{\cal A}$ such that $\pi\circ s = id$. If gauge
fixing is possible, we will be able to define a global section $s$ satisfying
these properties i.e. the corresponding fiber bundle $\tilde{\cal A}\rightarrow
\tilde{\cal A}/\cal G_{\ast}$ will be trivial. Then $\tilde{\cal
A}\approx \tilde{\cal A}/\cal G_{\ast} \times \cal G_{\ast}$. For
some positive or zero $q$
\begin{equation}
 \pi_{q}(\tilde{\cal A})\approx \pi_{q}(\tilde{\cal A}/{\cal G}_{\ast})
                              \otimes \pi_{q}(\cal G_{\ast})
\end{equation}
The space $\tilde{\cal A}$ is an affine space and as such it is contractible.
Therefore $\pi_{q}(\tilde{\cal A}) = 0$. So, if we are able to fix a gauge we
must have
\begin{equation}
0\approx \pi_{q}(\tilde{\cal A}/{\cal G}_{\ast})
                               \otimes \pi_{q}(\cal G_{\ast})
\end{equation}
In order for this condition to hold for any $q$, the homotopy groups of the
right-hand side should be vanishing. By examining in detail the homotopy groups
of $\cal G_{\ast}$ we will prove that this is not the case thus establishing
the impossibility of gauge fixing.\\

      \begin{center}
         \large\sc 3.\ \ Automorphism groups of principal fiber bundles\\ %%%%%
      \end{center}

\rm In this section we present the general formalism for determining the
conjugacy classes of the group of gauge transformations \cite{MP}. Assume that
$M$ is
a manifold, $P$ a principal $G-$ bundle over $M$ with the projection $\pi :
P\rightarrow M$. We denote the space of gauge transformations of this bundle by
${\cal G}_{\ast}(\pi)$. This notation indicates that we consider the base
manifold $M$ as well as the typical fiber $G$ fixed and we only vary the
projection $\pi$. In this way we are able to obtain all the possible $G-$
bundles over $M$. As we mentioned before, ${\cal G}_{\ast}(\pi)$ is a
topological group with topology induced from the compact-open topology of the
space $Map_{\ast}(P,P)$. Let $\{U_{\alpha}, \ \alpha\in I\}$ be a covering
of $M$ and $\{(U_{\alpha}, \phi_{\alpha}), \ \alpha\in I\}$ a chart. For
every $x\in U_{\alpha}$ let
$\phi_{\alpha x} : G\rightarrow \pi^{-1}(x) = P_{x}$
be the map $\phi_{\alpha x}(g) = \phi_{\alpha}(x,g)$. Define
$\theta_{U_\alpha}:{\cal G}(\pi_{\alpha}) \rightarrow Map_{\ast}(U_{\alpha},G)$
by the condition
\begin{equation}
\theta_{U_\alpha} = \phi_{\alpha x}^{-1}\circ f_{x}\circ \phi_{\alpha x}
\end{equation}
where $f$ denotes the right action of $G$ on $P$, $f_{x}$ the restriction of
$f$ to the fiber $P_x$ and $\pi_{\alpha}$ the restriction of $\pi$ to the chart
$U_{\alpha}\times G$. Locally every fiber bundle is trivial, so the map
$\theta_{U_\alpha}$ is an isomorphism of topological groups. Let's consider the
group
            $$\prod_{\alpha\in I} Map_{\ast}(U_{\alpha}, G)$$
and for every $\alpha\in I$ the restriction map $r_{\alpha} : {\cal
G}(\pi)\rightarrow {\cal G}(\pi_{\alpha})$. We define the function
\begin{equation}
\theta : {\cal G}(\pi)\rightarrow \prod_{\alpha\in I}
                                                     Map_{\ast}(U_{\alpha},G)
\end{equation}
with $\theta = \{\theta_{U_{\alpha}}\cdot r_{\alpha}\}_{\alpha\in I}$. \ \
$\theta$ is an embedding of topological groups which shows that the group
${\cal G}(\pi)$ corresponding to the fibration $P\stackrel{\pi}{\rightarrow}M$
can be considered as a subgroup of
               $$\prod_{\alpha\in I} Map_{\ast}(U_{\alpha},G)$$
${\cal G}(\pi)$ divides $\prod_{\alpha\in I} Map_{\ast}(U_{\alpha},G)$ into
conjugacy classes each of which is denoted by $\overline{{\cal G}(\pi)}$ and
$C(M,G)$ denotes the set of all conjugacy classes of the groups ${\cal G}(\pi)$
Let $k : M\rightarrow BG$ be a classifying map for $(P, \pi , M)$, $i$ be the
isomorphism $i : G/ZG \simeq In(G)$ where $In(G)$ is the group of inner
automorphisms of $G$ and $l$ the quotient homomorphism $l : G\rightarrow G/ZG$
(where $ZG$ denotes the center of $G$). Consider the map $\eta : G\rightarrow
In(G)$ defined by $\eta = i\circ l$ and let $B\eta : BG\rightarrow BIn(G)$ be
the induced map at the level of the classifying spaces $BG$ and $BIn(G)$.
Assume that $G$ has the form $A\times K$ where $A$ is a path-connected
Abelian group ($U(1)$ in the finite dimensional case) and $K$ any
path-connected group with trivial center. Let $(B\eta)_{\ast}$ be the map
\begin{equation}
(B\eta)_{\ast} : [M, BG]_{\ast}\rightarrow [M, BIn(G)]_{\ast}
\end{equation}
induced by $B\eta$ on the homotopy classes of based maps $Map_{\ast}(M,BG)$ and
$Map_{\ast}(M,BIn(G))$. If moreover $card [M, BIn(G)]_{\ast} = 1$ then the map
$\Xi : C(M,G)\rightarrow [M,BIn(G)]_{\ast}$ defined by
       $$\Xi (\overline{{\cal G}(\pi)}) = (B\eta)\circ k$$
is a bijection. The previous theorem establishes a very interesting
correspondence. Generally, it is very hard to compute $C(M,G)$ directly.
Instead, when the conditions of the theorem hold, we have to compute the
homotopy classes of maps $[M, BIn(G)]_{\ast}$ which is by comparison a much
easier space to handle. One very interesing consequence of the above theorem is
in the case in which $[M, BIn(G)]_{\ast}$ is trivial. Then $C(M,G)$ has only
one element. In that case the ``twisted" fiber bundle with projection map
$P\stackrel{\pi}{\rightarrow}M$ has an automorphism group which is conjugate to
that of the trivial fiber bundle $P\stackrel{\pi_0}{\rightarrow}M$. But it is
known that in the latter case the group of gauge transformations is the group
$Map_{\ast}(M,G)$. By that we have succeded in reducing the calculation of the
group of gauge transformations of the fiber bundle $\pi : P\rightarrow M$ to
the computation of the space $Map_{\ast}(M,G)$ which is much more manageable.
This is the strategy that we follow in the next section to compute the homotopy
type of the automorphism group of $SU(m)\times U(1)$.\\

      \begin{center}
         \large\sc 4.\ \ Application to $SU(m)\times U(1)$ over $S^n$\\
      \end{center}

\rm Let $G=SU(m)\times U(1)$ and $M=S^n$. We have the isomorphisms
   $$[S^{n}, B(SU(m)\times U(1))]_{\ast} \simeq [S^{n-1}, SU(m)\times
                                                                 U(1)]_{\ast}$$
By definition
   $$[S^{n-1}, SU(m)\times U(1)]_{\ast} \equiv \pi_{n-1}(SU(m)\times U(1))
                              \simeq \pi_{n-1}(SU(m))\otimes \pi_{n-1}(U(1))$$

\noindent For $n=3$ we get $\pi_{2}(SU(m))\simeq 0$\\
For $n=4$ we get $\pi_{3}(SU(m))\simeq Z$\\
Generally we can prove that $\pi_{n}(U(1)) = 0, \ \ \ \ n>1$. Taking this into
account
 $$[S^{n-1}, SU(m)\times U(1)]_{\ast} \simeq \pi_{n-1}(SU(m))$$
We also know that
 $$[S^{n}, BIn(SU(m)\times U(1))]_{\ast}\simeq [S^{n-1}, In(SU(m)\times
                                                               U(1))]_{\ast}$$
and this means that
$$[S^{n-1}, In(SU(m)\times U(1))]_{\ast} \simeq \pi_{n-1}(In(SU(m)\times
                                                                       U(1))$$
But
         $$Z(SU(m)\times U(1)) \simeq Z_{m}\times U(1)$$
Therefore
 $$In(SU(m)\times U(1)) \simeq (SU(m)\times U(1))/(Z_{m}\times U(1))$$
To compute
             $$\pi_{n}((SU(m)\times U(1))/(Z_{m}\times U(1)))$$
we observe that $SU(m)\times U(1)$ is a covering space for
$(SU(m)\times U(1))/(Z_{m}\times U(1))$. According to a theorem \cite{White}
when this is the case we get
 $$\pi_{n-1}((SU(m)\times U(1))/(Z_{m}\times U(1)))\otimes \pi_{n-1}(Z_m\times
                 U(1))\simeq \pi_{n-1}(SU(m)\times U(1)), \ \ \ n>3$$
With the compact-open topology $\pi_{n-1}(Z_m)\simeq 0$ and
$\pi_{n-1}(U(1))\simeq 0$ for these values of $n$.
Therefore
 $$\pi_{n-1}((SU(m)\times U(1))/(Z_{m}\times U(1)))\simeq
      \pi_{n-1}((SU(m)\times U(1)))\simeq \pi_{n-1}(SU(m))$$
Then we observe that the map
$$(B\eta)_{\ast} : [S^n, B(SU(m)\times U(1))]_{\ast}\rightarrow
                       [S^n, BIn(SU(m)\times U(1))]_{\ast}$$
is the trivial map so $card C(S^{n}, SU(m)\times U(1))=1$. This implies,
according to the theorem stated above, that ${\cal G}(\pi)$ is conjugate to
${\cal G}(\pi_{0})$ or that
\begin{equation}
{\cal G}(\pi) = Map_{\ast}(S^{n}, SU(m)\times U(1))
\end{equation}
Next we prove that the space $Map_{\ast}(S^{n},SU(m)\times
U(1))$ is not contractible. We begin
by noticing that in the compact-open topology
    $$Map_{\ast}(S^{n}, SU(m)\times U(1)) \approx Map_{\ast}(S^{n}, SU(m))
                                          \times Map_{\ast}(S^{n}, U(1))$$
where ``$\approx$" denotes homeomorphism. To reach our goal, it suffices to
prove that at least one homotopy group of $Map_{\ast}(S^n, SU(m)\times U(1))$
is
non-trivial. By applying the homotopy functor $\pi_{k}$ to the previous
equation we get
  $$\pi_{k}(Map_{\ast}(SU(m)\times U(1)) \simeq \pi_{k}(Map_{\ast}(S^n,SU(m))
        \times Map_{\ast}(S^n,U(1)))$$
{}From a known property of the homotopy functor for the Cartesian product of
topological spaces, this is equal to
\begin{equation}
\pi_{k}(Map_{\ast}(SU(m)\times U(1))) \simeq \pi_{k}(Map_{\ast}(S^n,SU(m)))
        \otimes \pi_{k}(Map_{\ast}(S^n,U(1)))
\end{equation}
By definition
 $$\pi_{k}(Map_{\ast}(S^n,U(1))) \equiv {[S^{k},
Map_{\ast}(S^n,U(1))]}_{\ast}$$
As a topological space $U(1)$ is $S^1$, therefore
 $$\pi_{k}(Map_{\ast}(S^n,U(1))) \equiv {[S^{k}, Map_{\ast}(S^n,S^1)]}_{\ast}$$
The topological spaces $S^n$, with the compact-open topology, are compact so
they are compactly generated. This does not guarantee that the spaces
$Map_{\ast}(S^n,S^1)$ equipped with the compact-open topology are also
compactly generated. In order to make them so, we retopologize
$Map_{\ast}(S^n,S^1)$ by applying to it the functor $k:Top_{\ast}\rightarrow
Comp$
where $Top_{\ast}$ is the category of pointed topological spaces and $Comp$ is
the category of compactly generated spaces. The functor $k$ is defined as
follows \cite{White} : let $X$ be a topological space. $k(X)$ and $X$ have the
same underlying
set. Let $A$ be a subset of $X$. Then $A$ is closed in $k(X)$ if and only if
$A\cap C$ is closed in $X$ for every compact subset $C$ of $X$. By
$Map_{\ast}(\ \ , \ \ )$ in the sequel we mean the element
$kMap_{\ast}(\ \ , \ \ )$ in $Comp$.
We also have by definition that
   $$[S^{k},Map_{\ast}(S^n,S^1)] \approx \pi_{0}(Map_{\ast}(S^{k},
                      Map_{\ast}(S^{n}, S^{1})))$$
In the category $Comp$ we know that
  $$Map_{\ast}(S^{k}, Map_{\ast}(S^{n}, S^{1})) \approx
                               Map_{\ast}(S^{k}\wedge S^{n}, S^{1})$$
where ``$\approx$" is a natural homeomorphism. The symbol ``$\wedge$" denotes
the smash product of the topological spaces $S^k$ and $S^n$. From the
definition of the smash product we see that
            $$S^{k}\wedge S^{n} \approx S^{k+n}$$
Therefore
   $${[S^k, Map_{\ast}(S^{n},S^{1})]}_{\ast} \simeq {[S^{k+n}, S^{1}]}_{\ast}
                                           \simeq \pi_{k+n}(S^{1})$$
Then
\begin{equation}
\pi_{k}(Map_{\ast}(S^{n}, U(1))) \simeq 0, \ \ \ \ \ k+n>1
\end{equation}
So, the second factor of the right-hand side of equation is trivial. In order
to compute the homotopy type of the first factor of equation we follow an
almost identical procedure. In this case we use the fact that for any
topological group $G$  the following homeomorhism is true
              $$G \simeq \Omega BG$$
where $BG$ is the classifying space of $G$ and $\Omega X$ denotes the loop
space of a topological space $X$. Remembering that, we have the following
equivalence
   $$Map_{\ast}(S^{n}, SU(m)) \simeq Map_{\ast}(S^{n}, \Omega BSU(m))$$
It is known though, that in the category of compactly generated spaces the
functors $\Sigma$ and $\Omega$ form an adjoint pair e.g. the following spaces
are naturally homeomorhic
 $$Map_{\ast}(S^{n}, \Omega BSU(m)) \simeq Map_{\ast}(\Sigma S^{n}, BSU(m))$$
where $\Sigma$ is the (reduced) suspension functor.
{}From the definition of the (reduced) suspension we can see that $\Sigma S^{n}
\approx S^{n+1}$.
Therefore
  $$Map_{\ast}(S^{n}, SU(m)) \simeq Map_{\ast}(S^{n+1}, BSU(m))$$
We want to calculate
  $$\pi_{k}(Map_{\ast}(S^{n+1}, SU(m)) \equiv
                                [S^{k}, Map_{\ast}(S^{n+1}, SU(m))]_{\ast}$$
Using the procedure described above for the case of $Map_{\ast}(S^{n}, U(1))$
we get
 $$\pi_{k}(Map_{\ast}(S^{n+1}, BSU(m)) \simeq {[S^{k+n+1}, BSU(m)]}_{\ast}
                                       \simeq \pi_{k+n+1}(BSU(m))$$
But from the definition of the loop functor $\Omega$ it follows that
 $$ \pi_{i}(SU(m)) \simeq \pi_{i}(\Omega BSU(m)) \simeq \pi_{i+1}(BSU(m))$$
So
\begin{equation}
\pi_{k}(Map_{\ast}(S^{n+1}, BSU(m))) \simeq \pi_{k+n}(SU(m))
\end{equation}
Combining equations (9), (10), (11) we find
\begin{equation}
\pi_{k}(Map_{\ast}(S^n, SU(m)\times U(1))) \simeq \pi_{k+n}(SU(m))
\end{equation}
This means that the space of automorhisms of the principal fiber bundle with
typical fiber $SU(m)\times U(1)$ over $S^n$ has the same homotopy groups as
$SU(m)$. The only difference is the shift of the index of the homotopy groups
of $SU(m)$ by $n$.\\

The physically interesting theory is the $SU(2)\times U(1)$ (the
Glashow-Salam-Weinberg model of the electroweak interactions) and potentially
an $SU(m)\times U(1)$ theory (probably as a byproduct of a grand unified
theory). The physically interesting dimensions of spacetimes are $n=3$
(corresponding to a one-point compactified spatial slice of a Lorentz manifold)
and $n=4$ (corresponding to the one-point compactification of $R^{4}$).
So for the physically interesting cases we have
\begin{equation}
\pi_{1}(Map_{\ast}(S^{3}, SU(2)\times U(1))) \simeq \pi_{4}(SU(2))\simeq Z_{2}
\end{equation}
\begin{equation}
\pi_{1}(Map_{\ast}(S^{4}, SU(2)\times U(1))) \simeq \pi_{5}(SU(2)) \simeq Z_{2}
\end{equation}
Therefore the spaces $Map_{\ast}(S^{3}, SU(2)\times U(1))$ and
$Map_{\ast}(S^{4}, SU(2)\times U(1))$ are not contractible. This means that any
$SU(2)\times U(1)$ theory over $S^3$ and $S^4$ has Gribov ambiguitites.\\

We also see that by using equation $(12)$ we can prove that many more theories
with gauge groups $SU(m)\times U(1)$ over $S^n$ have Gribov ambiguities. These
results also apply for gauge groups of the form $SU(m)\times [U(1)]^r$ which
are quite often encountered as gauge groups of superstring inspired models.\\

We can repeat the procedure given in this section for the group $SO(m)\times
U(1)$, $m$:odd, with similar results. All the arguments presented above for the
case of $SU(m)\times U(1)$ carry through in this case. The end result is that
   $$\pi_{k}(Map_{\ast}(S^{n}, SO(m)\times U(1))) \simeq \pi_{k+n}(SO(m))$$
Theories with gauge groups $SO(m)\times U(1)$ may conceivably arise as
intermediate steps  of the symmetry breaking pattern GUT $\rightarrow$ Standard
Model using supersymmetric or non-supersymmetric GUT as the starting point.
They may also arise as effective field theories for some systems or even as toy
models for testing new ideas. In all these cases we expect the index $m$ of the
gauge group $SO(m)$ to be relatively small (probably $m\leq 5$).\\

      \begin{center}
         \large\sc 5.\ \ Discussion and Conclusions\\
      \end{center}

\rm In this paper we proved that the Gribov ambiguity exists for $SU(m)\times
U(1)$  over the $S^n$. This is a
result which should not come as a big surprise, in view of the fact that the
non-Abelian gauge theories with semi-simple gauge group exhibit this
behavior. What is remarkable though, is that although we did not consider a
trivial fiber bundle over $S^n$, it turned out that the groups of gauge
transformations were conjugate to the ones of the corresponding trivial fiber
bundles. The spirit of this approach is strongly reminiscent of the approaches
encountered in Galois theory. There, we reduce a problem involving a
non-commutative group to a problem involving the automorphism group of that
group which is commutative. This simplifies the algebraic structure
considerably and allows us to obtain results that are unreachable through more
conventional methods. Similarly, in the present paper, instead of considering a
``twisted" principal fiber bundle, we analyse its automorphism group. It turns
out that this automorhism group is conjugate to that of a trivial fiber bundle.
Therefore many problems of the ``twisted" structure can be addressed in the
``untwisted" case. It would be hard to
imagine that this would be the case before solving the problem.\\

In section 2 we defined the moduli space $\tilde{\cal A}$. Using the same ideas
of framed connections we see that
  $${\cal G}/{\cal G}_{\ast} \approx Aut(P_{x_0}) \approx G$$
This statement as well as the lack of orbifold type singularities of $\tilde
{\cal A}$ and the existence of singularities for $\cal A$ may tempt us to
believe that the Gribov ambiguities have their source in the existence of
gauge transformations not continuously connected to the identity, in short in
the fact that $\pi_{0}({\cal G})\neq 0$. It has been known for some time
that this is not the case \cite{Baal}. Therefore potential connection between
global
anomalies and the Gribov ambiguity does not exist in that context so far as we
know today. Gribov copies exist even within the first Gribov horizon. It is not
yet known which is the maximal subspace of the moduli space $\tilde{\cal A}$
that does not have Gribov copies. A first step in the direction of the
determination of that fundamental modular domain is the proof that all gauge
orbits pass inside the Gribov horizon.\\

It would be more informative to extend our analysis to a base manifold that
does not have the topology of an $n-$sphere or of a product of spheres. This
can be, presumably, done by using a combination of the methods used in this
paper and a Postnikov system \cite{Bott},\cite{Jungman}
in which we can decompose the more complicated
base manifold. However, apart from its mathematical interest, we do not expect
this approach to give anything unexpected (i.e. the non-existence of a Gribov
ambiguity in the generic case).\\

The main task is to understand the physical meaning of the Gribov
ambiguity and the effects that it has, not only on the formal structure of
non-Abelian gauge theories but also on observable quantities. The former case
will provide a better understanding of the non-perturbative structure of gauge
theories. The
latter goal can also be attained, in principle. It is our opinion, however,
that the Gribov ambiguity is most likely a gauge artifact, therefore it should
not be expected to have any
observable consequences. It may present considerable complication in the
calculations,in handling global issues, but the lack of its existence in
algebraic gauges, seems compelling enough to make us suspect that it is a
mathematical artifact. We believe that the issue can be settled after
performing a BRST analysis and see the form that the Gribov ambiguity takes in
that formalism.\\

\noindent\underline{Acknowledgement} \ \ \
I am grateful to Gaunce Lewis Jr. for a discussion in homotopy theory and to
Mark Bowick for his comments on the manuscript.\\

\end{document}